\documentclass[aip,reprint]{revtex4-1}

\pdfoutput=1

\draft 

\usepackage{graphicx}
\usepackage{dcolumn}
\usepackage{bm}

\begin{document}


\title{Ionic-liquid-gating setup for stable measurements and reduced electronic inhomogeneity at low temperatures} 



\author{Yamaguchi Takahide}
\affiliation{International Center for Materials Nanoarchitectonics, National Institute for Materials Science, Tsukuba 305-0044, Japan}
\affiliation{University of Tsukuba, Tsukuba, 305-8571, Japan}

\author{Yosuke Sasama}
\affiliation{International Center for Materials Nanoarchitectonics, National Institute for Materials Science, Tsukuba 305-0044, Japan}
\affiliation{University of Tsukuba, Tsukuba, 305-8571, Japan}

\author{Hiroyuki Takeya}
\affiliation{International Center for Materials Nanoarchitectonics, National Institute for Materials Science, Tsukuba 305-0044, Japan}

\author{Yoshihiko Takano}
\affiliation{International Center for Materials Nanoarchitectonics, National Institute for Materials Science, Tsukuba 305-0044, Japan}
\affiliation{University of Tsukuba, Tsukuba, 305-8571, Japan}

\author{Taisuke Kageura}
\affiliation{Waseda University, Tokyo 169-8555, Japan}

\author{Hiroshi Kawarada}
\affiliation{Waseda University, Tokyo 169-8555, Japan}


\date{\today}

\begin{abstract}
The ionic-liquid-gating technique can be applied to the search for novel physical phenomena at low temperatures because of its wide controllability of the charge carrier density. Ionic-liquid-gated field-effect transistors are often fragile upon cooling, however, because of the large difference between the thermal expansion coefficients of frozen ionic liquids and solid target materials. In this paper we provide a practical technique for setting up ionic-liquid-gated field-effect transistors for low-temperature measurements. It allows stable measurements and reduces the electronic inhomogeneity by reducing the shear strain generated in frozen ionic liquid.
\end{abstract}


\maketitle 


\section{Introduction}

The application of the ionic-liquid-gating technique to low-temperature physics has attracted considerable attention because it can control the charge carrier density over an extremely wide range.\cite{Bis17} This technique uses an ionic liquid (organic salt in the liquid phase at room temperature) as the gate dielectric in field-effect transistors in which a target material acts as a channel of charge carriers. In this electric double layer transistor (EDLT), the large capacitance of the electric double layer at the surface of the target material allows the channel to have a large concentration of charge carriers on the order of $10^{13}-10^{14}$ cm$^{-2}$. Even when the EDLT is cooled to below the freezing point of the ionic liquid, the electric double layer and the resulting charge carrier density are preserved. Therefore, one can measure the low-temperature electronic properties of the sample with a charge carrier density controlled in a wide range. Electric-field-induced superconductivity has been obtained in various materials by using ionic-liquid gating.\cite{Ye10,Bol11,Uen11,Len11,Dub12,Ye12,Jo15,Shi15,Sai15,Lu15,Li16,Cos16,Zen18}

\begin{figure}
\includegraphics[width=7truecm]{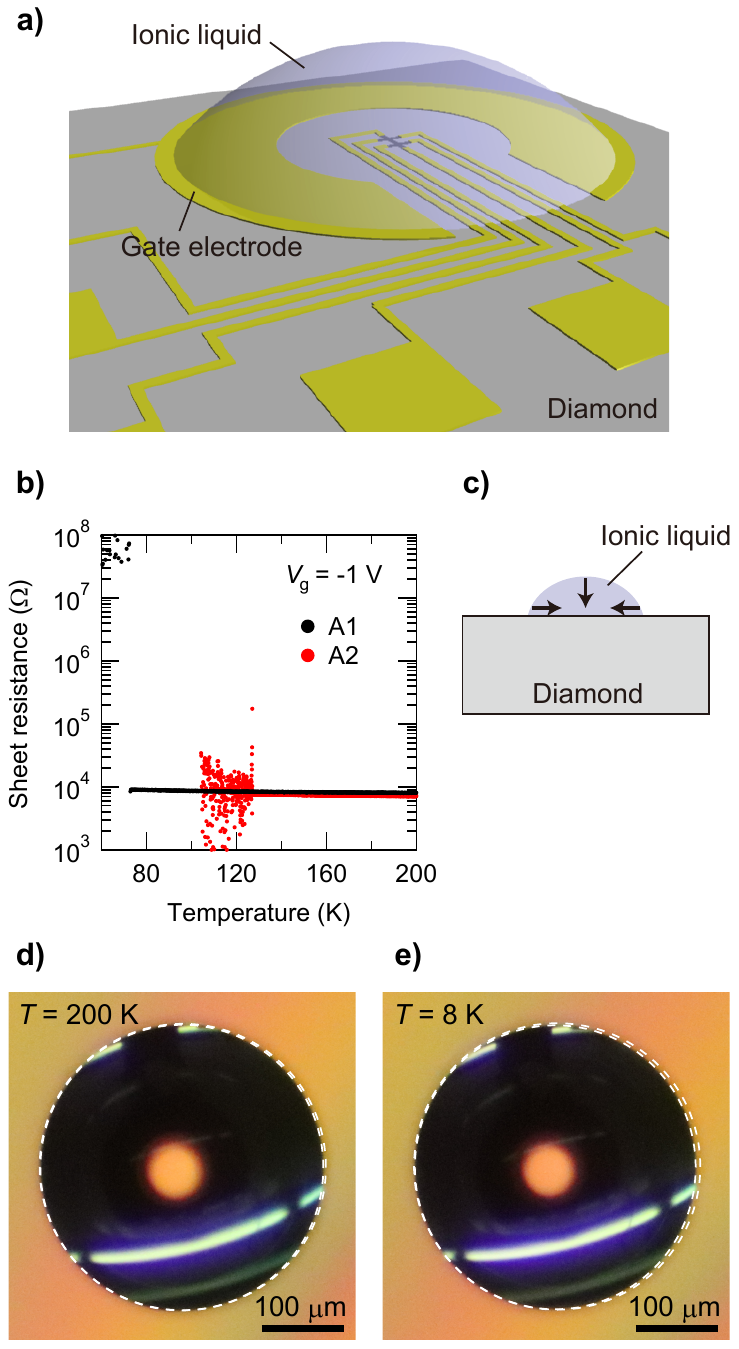}
\caption{(a) Schematic of a diamond EDLT with no structure for reducing the shear strain in the ionic liquid. Drain, source, and gate electrodes have been fabricated on the surface of a single crystal diamond. Only the channel of the EDLT on the diamond surface is hydrogen-terminated, which assists the accumulation of holes. The rest of the surface is oxygen-terminated. The oxygen termination electrically isolates the channel from the gate. A drop of ionic liquid is applied to cover both the channel and gate electrode. (b) Temperature dependence of the sheet resistance of two diamond EDLTs at the gate voltage of $-1$ V. The sample A1 was set the same as shown in (a) and measured with a two-point configuration. The ionic liquid on the sample A2 was covered with an unfixed small piece of glass. The sample A2 was measured with a four-point configuration. The resistance of A1 suddenly increased (the current decreased to the noise level) at 73 K, which was caused by the fracture of ionic liquid. The source and drain electrodes on the hydrogen-terminated diamond surface were also partially peeled off. Similar destruction of the device structure also occurred for A2 at 127 K. (c) The large thermal expansion coefficient of frozen ionic liquid compared to that of diamond leads to shear strain (and resulting slips) at the interface. (d,e) Optical microscope images of a droplet of ionic liquid (DEME-BF$_4$) on a hydrogen-terminated diamond surface at 200 K (d) and 8 K (e). The contraction of the frozen ionic liquid was observed, presumably because the hydrogen-terminated surface of diamond is highly "non-stick". The droplet size at each temperature and at 300 K is depicted by dashed lines.}
\end{figure}

However, there is a practical problem for such low-temperature measurements on EDLTs: frozen ionic liquids often fracture at low temperatures. This induces some detrimental effects on measurements, such as a sudden jump in the resistance-temperature curve (Fig. 1). A large electronic inhomogeneity possibly due to the local detachment of frozen ionic liquid from the sample has also been reported for WS$_2$ and MoS$_2$ EDLTs.\cite{Jo15,Cos16} These problems are presumably due to the shear strain caused by the large difference in the thermal expansion coefficient between the frozen ionic liquid and target sample or its substrate. 

In this paper we introduce an experimental technique that suppresses the shear strain and leads to stable measurements on EDLTs at low temperatures. Our setups for diamond and silicon EDLTs\cite{Yam13,Yam14,Yam16,Sas17,Sas172} are shown as examples. This technique will allow stable and efficient low-temperature experiments with EDLTs and studies of high-quality samples with reduced electronic inhomogeneity.

\section{Results and discussion}

A key feature of our setup is a counter plate placed above the sample/substrate surface (Fig. 2). Ionic liquid is inserted between the counter plate and sample/substrate surface. A similar setup has been used in previous experiments.\cite{Bol11,Dub12} We propose here that an adequate spacing between the sample/substrate surface and the counter plate can reduce the shear strain that appears when the device is cooled. Our idea is to compensate for the cooling-induced shrinkage of frozen ionic liquid in a mechanical manner by using the shrinkage of the counter plate support. If the shrinkage of the support along the z axis is larger than that of the ionic liquid, the ionic liquid is compressed along the z axis and expands along the xy plane. If the shrinkage of ionic liquid along the xy plane due to cooling cancels this expansion, there should be no shear strain along the xy plane. The counter plate can be used as a gate electrode if its surface is electrically conductive and a proper wiring is made. 

\begin{figure}
\includegraphics[width=7truecm]{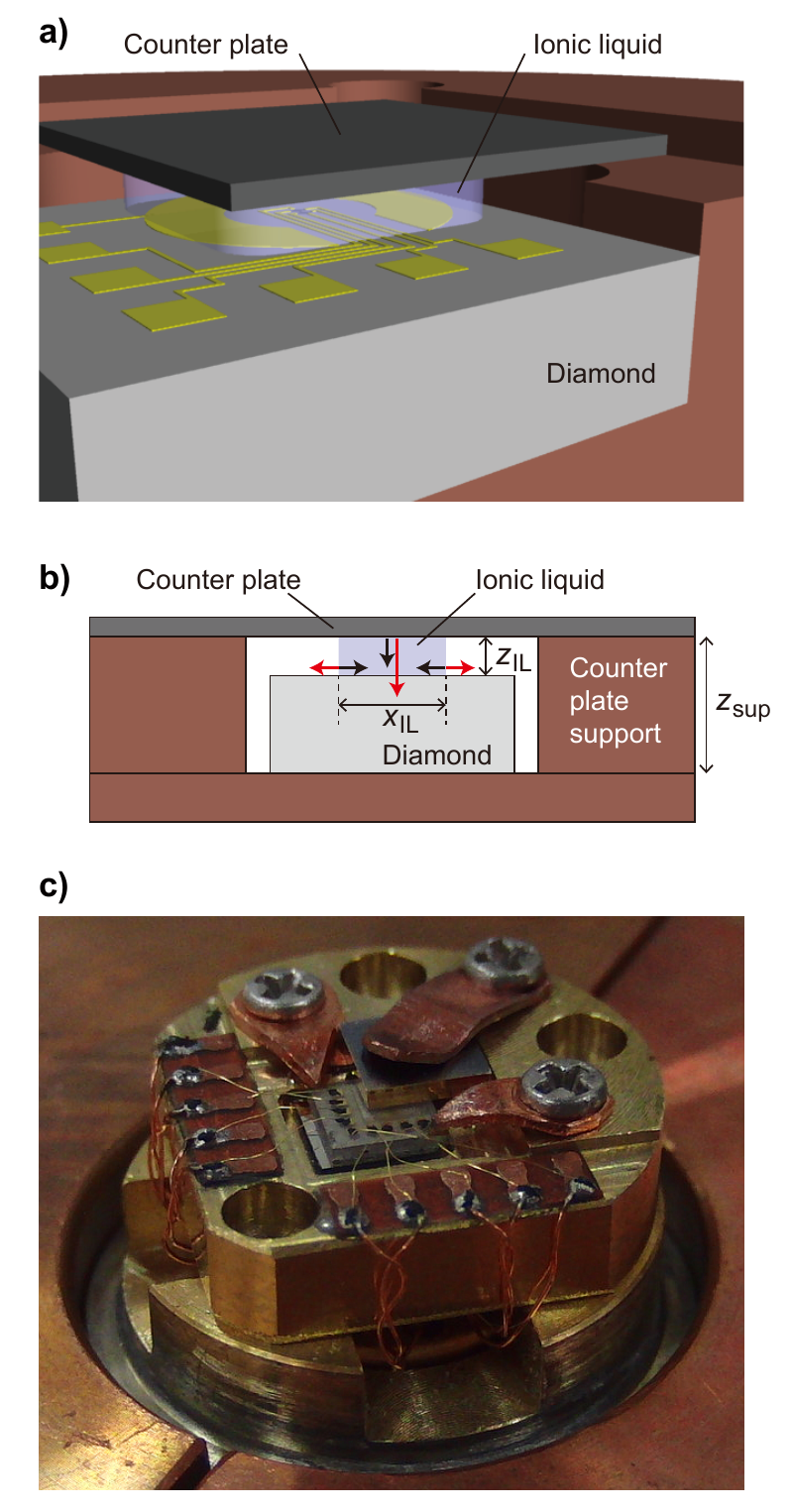}
\caption{(a,b) Schematic of a diamond EDLT with a counter plate. (c) Optical image of a diamond EDLT set on a sample holder with a counter plate. The sample holder is shown here on a holder stand, which is removed when the sample holder is sealed with indium (Fig. 3). The dimensions of the diamond sample are 2.6 mm $\times$ 2.6 mm $\times$ 0.3 mm.}
\end{figure}

\begin{figure}
\includegraphics[width=4.6truecm]{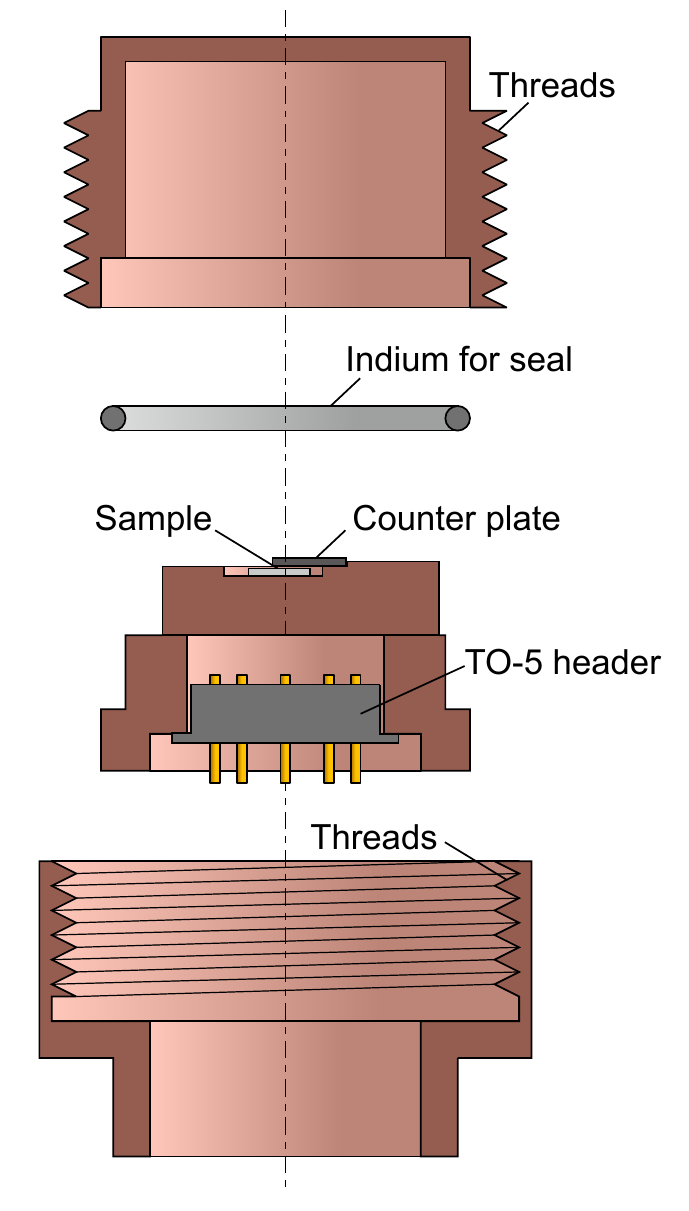}
\caption{Schematic of the sample holder assembly for sealing in a glove box. A TO-5 header is soldered to the main part of the sample holder for electrical feedthrough. }
\end{figure}

\begin{figure}
\includegraphics[width=6.5truecm]{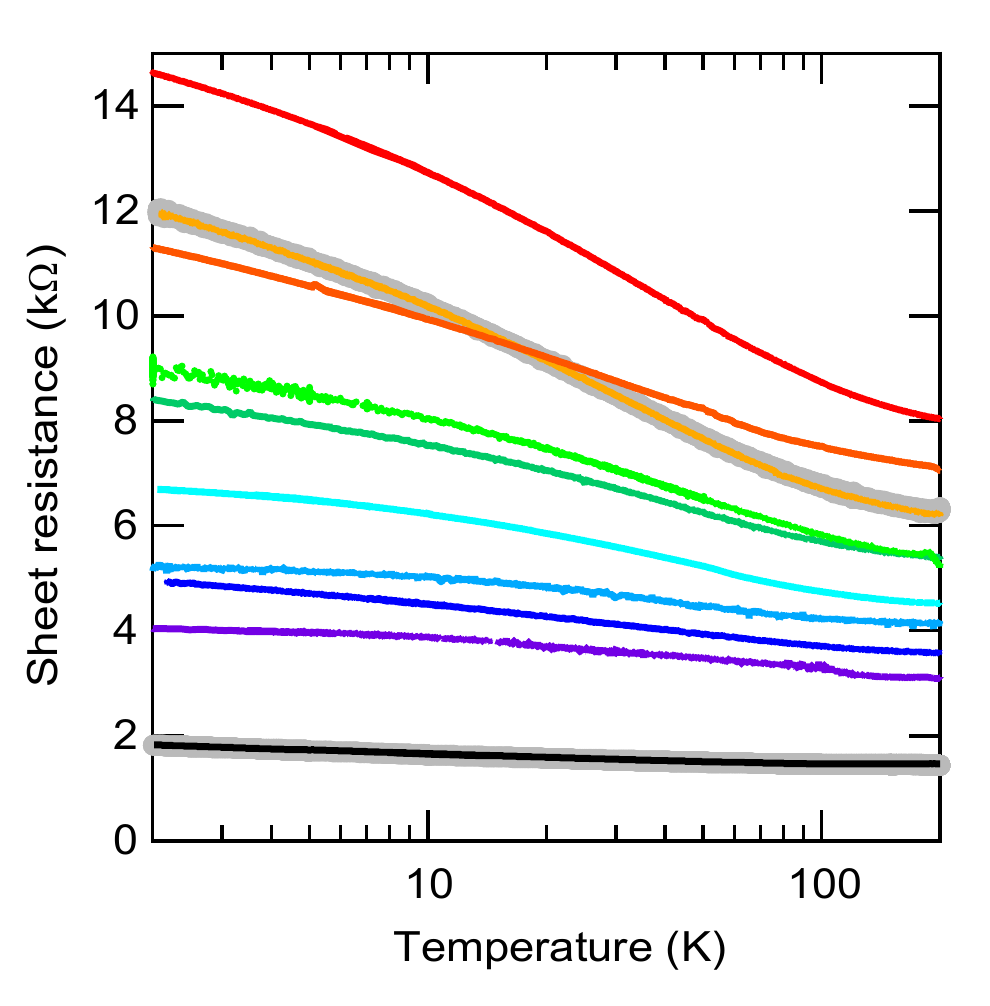}
\caption{Temperature dependence of the sheet resistance of 10 different diamond EDLTs. The resistance was measured with a four-point configuration. From top to bottom at the lowest temperature, the samples are numbered as B1-10. The surface orientation, ionic liquid, and gate voltage are as follows. B1: (100), DEME-TFSI, $-1.8 V$; B2: (100), DEME-BF$_4$, $-1.8 V$; B3: (111), DEME-BF$_4$; $-1.8$ V, B4: (100), DEME-TFSI, $-1.49$ V; B5: (100), DEME-BF$_4$, $-1.44$ V; B6: (100), TMPA-TFSI + HTFSI, $-2.58$ V; B7: (100), DEME-BF$_4$, $-1.96$ V; B8: (111), DEME-BF$_4$, $-2.4$ V; B9: (111), DEME-BF$_4$, $-2.2$ V; B10: (111), DEME-BF$_4$, $-1.8$ V. Boron-doped diamond was used as source, drain, and gate electrodes for B6. The diamond surface of B10 is atomically flat.\cite{Yam14} The curves measured while the sample was cooled down and warmed up are both shown for B2 and B10. (Gray thick lines are for warming up.) The resistance variation for different samples despite similar gate voltages is attributed to different amounts of charged adsorbates on the diamond surface.\cite{Yam13}}
\end{figure}

Let us examine the adequate spacing between the sample/substrate surface and the counter plate. We assume that the counter plate and its support are thick and rigid enough that they are not deformed by external force. We also assume that the thermal expansion coefficient of the sample (or substrate) is small and can be neglected. This is the case when diamond or silicon is used as a substrate or a sample, because at temperatures below 293 K the thermal expansion coefficients for diamond\cite{Sto11} and silicon\cite{Mid15} are less than $1{\times}10^{-6}$ and $3{\times}10^{-6}$ (K$^{-1}$), respectively, which are smaller than those of most of other materials. This assumption is only for simplification of the calculation shown below. With a straight forward modification, our scheme can be applied to any material.

The length variations of the frozen ionic liquid along the x direction due to the temperature change ${\Delta}T$ and along the z direction due to the mechanical force are given by
\begin{eqnarray}
\frac{{\Delta}x_\mathrm{IL}}{x_\mathrm{IL}}=\alpha_\mathrm{IL}{\Delta}T-{\sigma}_\mathrm{IL}\frac{{\Delta}z_\mathrm{IL}}{z_\mathrm{IL}},\\
{\Delta}z_\mathrm{IL}=-z_\mathrm{IL}\alpha_\mathrm{IL}{\Delta}T+z_\mathrm{sup}\alpha_\mathrm{sup}{\Delta}T.
\end{eqnarray}
Here ${\alpha}_\mathrm{IL}$ and ${\alpha}_\mathrm{sup}$ are the thermal expansion coefficients of the frozen ionic liquid and the counter plate support, and ${\sigma}_\mathrm{IL}$ is the Poisson's ratio of the frozen ionic liquid.
For the shear strain caused by the temperature change ${\Delta}T$ to be minimized, 
\begin{eqnarray}
\frac{{\Delta}x_\mathrm{IL}}{x_\mathrm{IL}}=0.
\end{eqnarray}
Then,
\begin{eqnarray}
z_\mathrm{IL}=\frac{{\sigma}_\mathrm{IL}}{1+{\sigma}_\mathrm{IL}}\frac{\alpha_\mathrm{sup}}{\alpha_\mathrm{IL}}z_\mathrm{sup}.
\end{eqnarray}
The value of Poisson's ratio $\sigma$ for most materials is $0.3-0.4$. We use brass and copper for the support of the counter plate. The thermal expansion coefficient of copper is $10{\times}10^{-6}$ (K$^{-1}$) at 100 K and $15{\times}10^{-6}$ (K$^{-1}$) at 200 K.\cite{Nix41} It is difficult to find the data of thermal expansion of frozen ionic liquids at temperatures below their freezing point. We directly observed the thermal contraction of droplets of ionic liquid (DEME-BF$_4$; freezing point: 238 K, melting point: 282 K\cite{Kim05}) on a hydrogen-terminated diamond surface under an optical microscope [Figs. 1(d) and 1(e)]. The diameter of the droplets shrank by $0.7{\pm}0.1{\%}$ when the temperature decreased from 200 to 8 K. If we assume that the thermal expansion coefficient of the frozen ionic liquid depends linearly on temperature, it is estimated to be $(35{\pm}5){\times}10^{-6}$ (K$^{-1}$) at 100 K and $(70{\pm}10){\times}10^{-6}$ (K$^{-1}$) at 200 K, which we use in the following calculation. These values are in the same range as those of organic charge-transfer salts, which are $(40-80){\times}10^{-6}$ (K$^{-1}$) at 100 K and $(40-80){\times}10^{-6}$ (K$^{-1}$) at 200 K.\cite{Mul02,Sou08,Fou13} The height $z_\mathrm{sup}$ of the support of the counter plate is $0.45-0.5$ mm in our experimental setup for diamond EDLTs. If we use these values, the thickness $z_\mathrm{IL}$ of ionic liquid should be $30-50$ $\mu$m for 100 K and $20-40$ $\mu$m for 200 K to minimize the shear strain. If a softer material with a larger $\alpha_\mathrm{sup}$ is used for the support (for example, polymer), then it is better to increase the ratio $z_\mathrm{IL}/z_\mathrm{sup}$. If the sample/substrate is fixed on the sample holder using adhesive tape, its large thermal expansion coefficient should also be taken into consideration. 

An optical microscope image of our setup for a diamond EDLT is shown in Fig. 2(c). The diamond is fixed using two copper claws, without the use of adhesive tape. As a counter plate, we used a Ti/Pt or Ti/Au deposited glass (or silicon) plate or a diamond substrate with a boron-doped layer on the surface. Here the counter plate also acted as a gate electrode. The thickness of the diamond differed from sample to sample because the original thickness of the diamond substrate and the amount of surface polishing differed. We adjusted the spacing between a sample and counter plate to be ${\approx}20-30$ $\mu$m each time by using some metal spacer plates with different thicknesses: 20, 30, 40, 50, 80, and 100 $\mu$m. This sample holder was also designed so that it can be sealed with indium in an Ar-filled glove box\cite{Bon00,Bon01,Brass} to prevent water contamination of the ionic liquid (Fig. 3). 

We have not observed any significant jumps in the temperature dependence of resistance of diamond EDLTs and could perform stable measurements at low temperatures with this setup. The temperature dependence of the resistances of ten different diamond EDLTs is shown in Fig. 4. The curves vary in a monotonic manner, although a few curves cross, possibly due to the difference in the surface crystallographic orientation. Furthermore, there is almost no difference between the resistance-temperature curves measured while the sample is cooled down and warmed up. This indicates that the local detachment of ionic liquid\cite{Jo15} is negligible during the thermal process. We observed an electric-field-induced insulator-metal transition and Shubnikov de-Haas oscillations of diamond with this setup.\cite{Yam13,Yam14} An anomalous low-temperature magnetotransport of the electric-field-induced charge carriers was also observed in diamond with the (100) surface.\cite{Yam16} 

\begin{figure}
\includegraphics[width=7truecm]{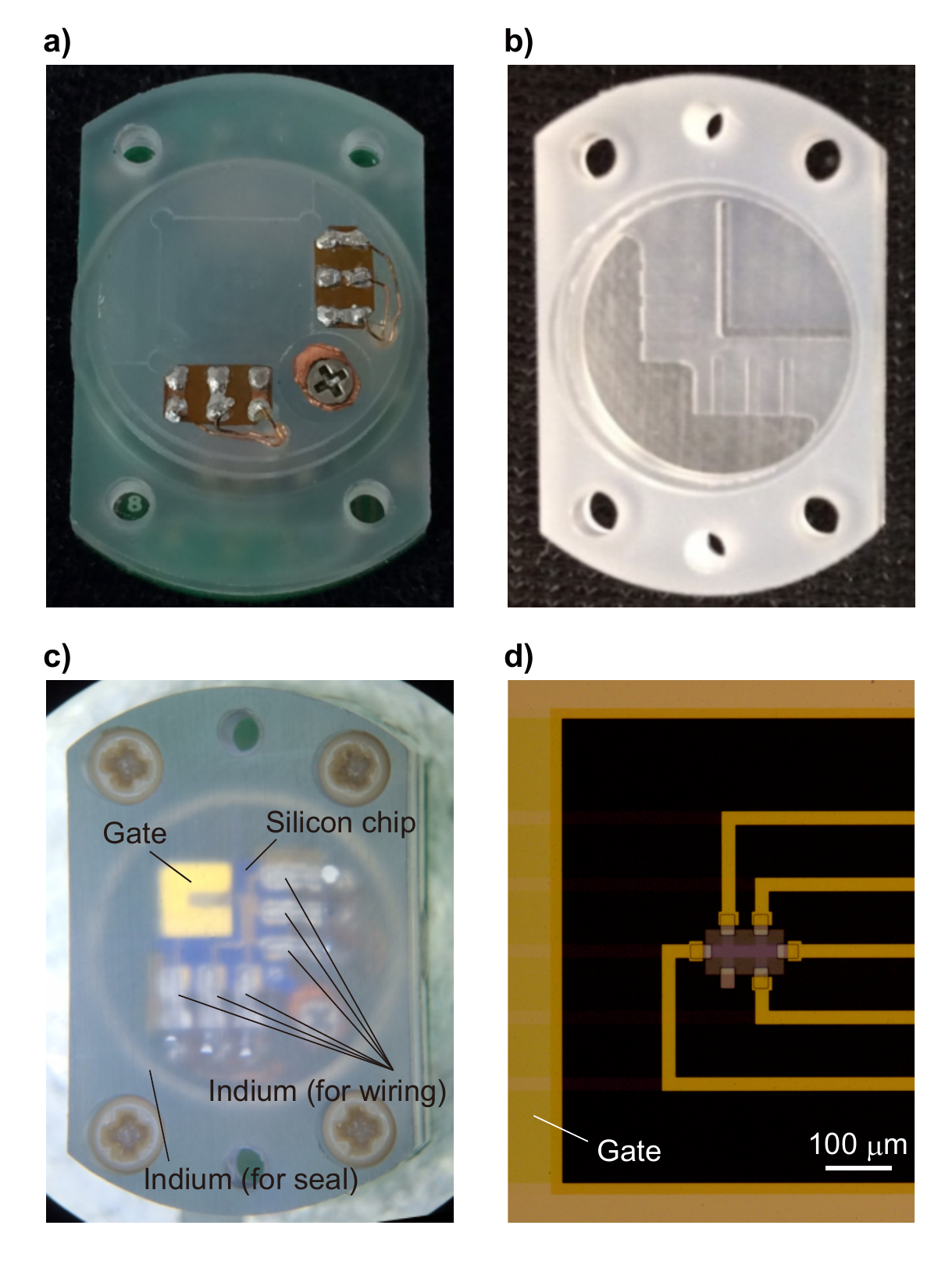}
\caption{Optical image of a sample holder for silicon EDLTs. (a) The main part of the sample holder. (b) The lid of the sample holder. (c) A silicon EDLT ready for low-temperature measurements. First, small pieces of indium for electrical wiring and the seal of the sample holder are placed on the lid. Then, a silicon chip with a hydrogen-terminated channel, Hall bar electrodes, and a gate electrode\cite{Sas17,Sas172} is fixed on the main part of the sample holder by a copper claw in an Ar-filled glove box. After a drop of ionic liquid is applied, the lid is screwed on. This makes the electrical wiring, the seal of the sample holder, and the insertion of the ionic liquid between the silicon and counter plate (lid) at the same time. The dimensions of the silicon chip are approximately 6.0 mm $\times$ 6.0 mm $\times$ 0.38 mm. (d) Optical image of the Hall bar and gate electrode on the silicon chip.}
\end{figure}

\begin{figure}
\includegraphics[width=8.2truecm]{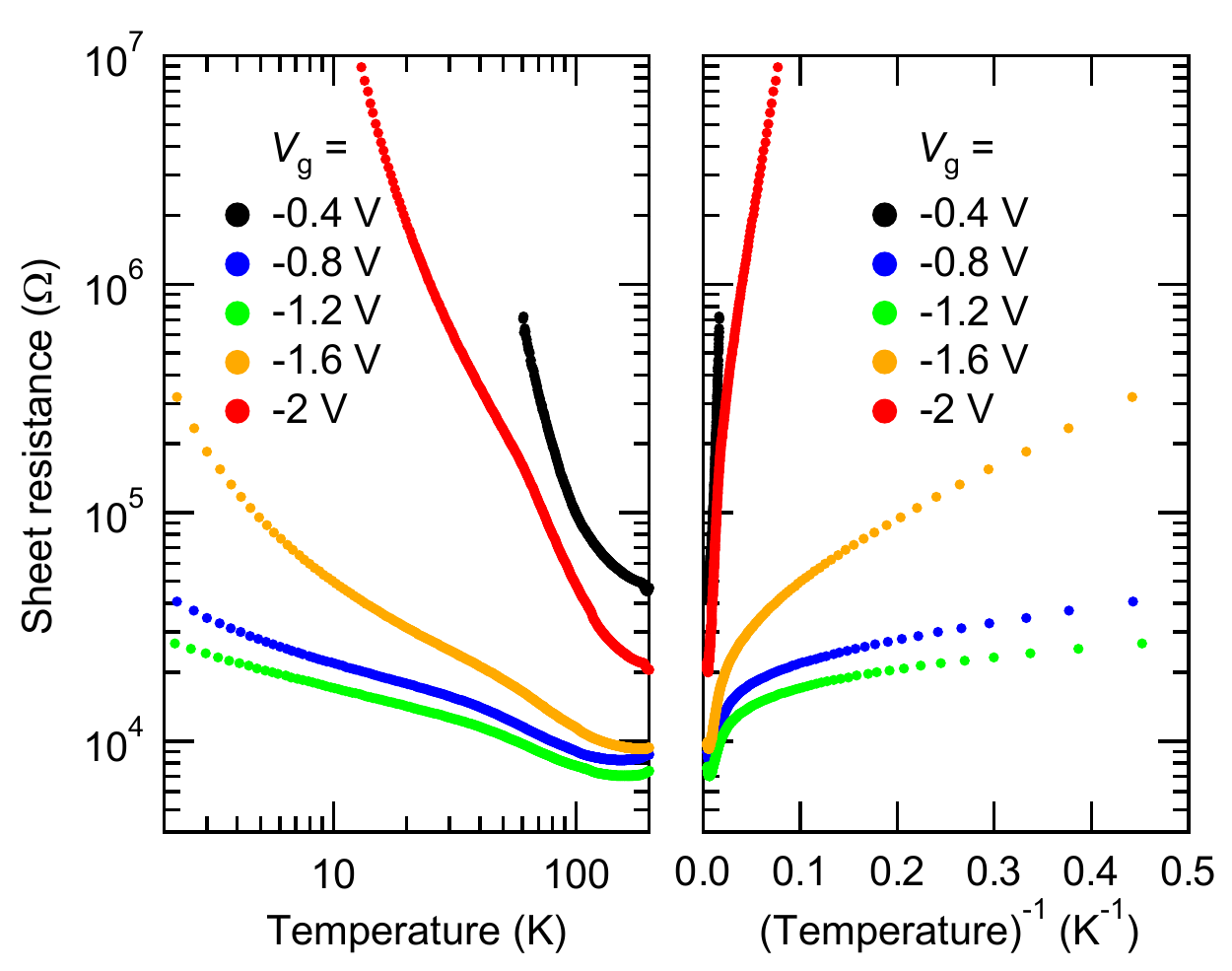}
\caption{Temperature dependence of the sheet resistance of a silicon EDLT for different gate voltages. The resistance was measured with a four-point configuration. The resistance could not be measured accurately at $T{\le}220$ and 60 K for $V_g=0$ and $-0.4$ V because the contact resistance became very high. The resistance decreased with increasing negative gate voltage at $V_g{\ge}-1.2$ V, but it increased at $V_g{\le}-1.2$ V. See Ref. 19 for details.}
\end{figure}

We performed a study of silicon EDLTs as well\cite{Sas17,Sas172}. Another type of sample holder (Fig. 5) was fabricated for the silicon EDLTs for the following reasons. The silicon surface of the channel of the EDLTs is hydrogen-terminated to reduce the trap density. This hydrogen termination is crucial for the device operation\cite{Sas17,Sas172} but, unlike the hydrogen termination of diamond surface, is easily destroyed by air exposure. Therefore, the electrical wiring between the sample and sample holder cannot be performed in air for the silicon EDLTs. The sample holder is designed so that the electrical wiring can be performed using small pieces of indium in an Ar-filled glove box. The sample holder can also be sealed with indium in the glove box. 

This sample holder is made of polychlorotrifluoroethylene (PCTFE) and the lid acts as a counter plate. The counter plate support consists of $\approx$0.40 mm thick PCTFE and $\approx0.1-0.15$ mm thick indium: $z_\mathrm{sup}{\approx}0.50-0.55$ mm. The thermal expansion coefficient of PCTFE is $34{\times}10^{-6}$ (K$^{-1}$) at 100 K and $47{\times}10^{-6}$ (K$^{-1}$) at 200 K.\cite{PCTFE} The coefficient ($\alpha_v/3$) for indium is $27{\times}10^{-6}$ (K$^{-1}$) at 100 K and $28{\times}10^{-6}$ (K$^{-1}$) at 200 K.\cite{Smi64} Using Eq. 4, $z_\mathrm{IL}$ for the minimized shear strain is estimated to be $100-170$ $\mu$m for 100 K and $60-110$ $\mu$m for 200 K. We set $z_\mathrm{IL}\approx120-170$ $\mu$m in the actual setup. By using this setup and using ion implantation underneath the electrodes to reduce the contact resistance, we were able to measure detailed low-temperature transport properties of silicon EDLTs\cite{Sas172} (Fig. 6).

The proposed method minimizes the shear strain in frozen ionic liquid, but the perfect elimination of this strain in an entire temperature range is difficult. This is because the temperature dependences of $\alpha_\mathrm{IL}$ and $\alpha_\mathrm{sup}$ generally differ. There may still be small remaining strains at low temperatures. The Shubnikov-de Haas oscillations observed in diamond EDLTs suggest a spatial inhomogeneity of charge carrier density and mobility at low temperatures.\cite{Yam14} Further work is necessary to elucidate whether this inhomogeneity has an intrinsic origin\cite{DezArXiv} or is caused by local distortion of the frozen ionic liquid due to residual shear strains. Detailed measurements of the thermal expansion coefficient and Poisson's ratio of ionic liquids at different temperatures are also awaited. It may be possible to further reduce shear strain by setting the spacing $z_\mathrm{IL}$ so that the integral of $(1/x_\mathrm{IL})(\mathrm{d}x_\mathrm{IL}/\mathrm{d}T)$ (dependent of $\alpha_\mathrm{IL}(T)$, $\sigma_\mathrm{IL}(T)$, and $\alpha_\mathrm{sup}(T)$) between the temperature of interest and the freezing temperature of the ionic liquid would be zero.

\section{Conclusions}

We proposed a practical method to reduce shear strain in frozen ionic liquid for stable measurements of electric double layer transistors at low temperatures. The reduction of shear strain was achieved by compensating for the cooling-induced shrinkage of frozen ionic liquid in a mechanical way using a counter plate and its support. The simple setup will be used for various materials and allow stable and efficient experiments at low temperatures. In particular, it prevents frozen ionic liquid from detaching from the sample surface and thus prevents the device breakdown due to cooling. It will also reduce the electronic inhomogeneity caused by the shear strain and thus help to study more intrinsic properties of the target materials.

\section{Acknowledgments}

We appreciate helpful comments from Y. Ootuka and thank E. Watanabe, H. Osato, D. Tsuya, S. Hamada and S. Tanigawa for device fabrication in the early stage of this study. This study was supported by Grants-in-Aid for Fundamental Research (Grant Nos. 25287093 and 26220903) and the "Nanotechnology Platform Project" of MEXT, Japan.


%
%

%


\bibliography{EDLTsetupbib}

\begin{thebibliography}{32}
\bibitem [{Bra()}]{Brass}%
  \BibitemOpen
  \href@noop {} {}\bibinfo {note} {This sample holder was made of brass
  following Ref. 27. We noticed that this brass showed a weak superconductivity
  at temperatures below 7.4 K, probably due to its lead impurities. In
  addition, the TO-5 header for electrical feedthrough is made of a
  ferromagnetic material. Using a Hall sensor, we confirmed that the influence
  of these on the magnetic field applied to the sample was negligible at
  $2-300$ K. This is consistent with experiments in Refs. 27 and
  28.}\BibitemShut {Stop}%
\end{thebibliography}

\end{document}